\documentstyle[preprint,aps]{revtex}
\begin{document}
\draft
\title{
$\pi$-$\eta$ MIXING FROM QCD SUM RULES}
\author{Chuan-Tsung Chan$^{(1)}$, E. M. Henley$^{(1),(2)}$, and
Th. Meissner$^{(2)}$}
\address{$^{(1)}$ Physics Department, FM-15, University of Washington,
Seattle, WA 98195 \\
$^{(2)}$ Institute for Nuclear Theory, HN-12, University of Washington,
Seattle, WA 98195}
\date{March, 1994}
\maketitle
\begin{abstract}
The $q^2$ dependence of the $\pi$-$\eta$ mixing amplitude is examined with
the use of QCD sum rules.  The linear slope of the mixing function
$\theta(q^2)$ is found to be much smaller than that
for $\rho$-$\omega$ mixing.  Thus the mixing
amplitude is approximately the same in the space-like region as in the
time-like one, and one may neglect the $q^2$ dependence of the mixing.
A comparison between a hadron-meson, an effective chiral model, and the QCD
sum rules approaches is made.
\end{abstract}
\newpage
\section{INTRODUCTION}

The recent interest in examining the off-shell behavior of $\rho$-$\omega$
mixing \cite{1,2,3,4}
and $\pi$-$\eta$ mixing \cite{5,6,7}
is based on the observation of
Goldman, Henderson and Thomas (GHT) \cite{1}
that the meson mixing previously
used in the calculation of the nucleon-nucleon (N-N) charge-symmetry-violating
(CSV) potential \cite{8}
may not be correct.  In particular, the value of
$\rho$-$\omega$ mixing found experimentally in the time-like region $(q^2 >
0)$ may not apply in the space-like region, from which the CSV potential is
generated.  Subsequently, other authors have discussed this issue for
$\rho$-$\omega$ mixing by using different effective models \cite{1,3,4}, and
also find appreciable $q^2$ dependence of the mixing amplitude.  With the
advent of QCD sum rules \cite{9} and their phenomenological success in
describing hadronic properties \cite{10,11,12},
it is natural to apply this technique
to meson-mixing, particularly since the method is closer to QCD than other
models that have been used.

In the pioneering work of Shifman, Vainshtein and Zakharov (SVZ) \cite{9},
the ``on shell" $\rho$-$\omega$ mixing was used as an illustration of the
power of QCD sum rules.  However, in their paper some fine points remain
elusive and there is no discussion of the ``off-shell" behavior of
meson-mixing.  These details were re-examined by Hatsuda, Henley, Meissner
and Krein (HHMK) \cite{4}, who used both Borel and finite-energy sum rules,
together with dispersion relations to determine the momentum-dependence of
$\rho$-$\omega$ mixing.  They found a rapid $q^2$ variation of the mixing
parameter $\theta(q^2)$.  In this paper we shall follow the spirit of HHMK
in applying it to $\pi$-$\eta$ mixing.  By performing the Borel analysis,
we can extract
$\theta(q^2)$ as a function of the CSV parameters,
and we can also study
how the QCD condensates affect the mixing function $\theta(q^2)$.

This paper is organized as follows: In section I, we establish notations and
definitions.  The calculations are described in section III.  In section IV,
we relate our results to those of GHT \cite{1} and Piekarewicz-Williams (PW)
\cite{3}.  Finally, in section V, conclusions and a summary are presented.

\section{$\pi$-$\eta$ MIXING (FORMALISM)}

The $\pi$-$\eta$ mixing function $\theta(q^2)$ is defined by the mixed
correlator of $\pi^0$ and $\eta^0$
\begin{equation}
\pi^{\pi\eta}(q^2) = i\int d^4x\;e^{iqx}<T\pi^0(x)\eta^0(0)>
\end{equation}
where $\pi^0$ and $\eta^0$ are eigenstates of an isosymmetric
strong-interaction Hamiltonian.

We can express the physical $\pi$-$\eta$ fields, which are eigenstates of
the full strong interaction Hamiltonian, in terms of $\pi^0$, $\eta^0$, by
introducing the mixing matrix
\begin{equation}
\left(\begin{array}{c}
\pi \\
\eta
\end{array}\right) = \left(\begin{array}{lr}
                     1&\epsilon \\
                     -\epsilon & 1
                     \end{array}\right)
\left(\begin{array}{c}
      \pi^0 \\
      \eta^0
      \end{array}\right),
\end{equation}
where $\epsilon$ is a ``mixing angle".  By saturating the
$Im\;\pi^{\pi\eta}$ with the $\pi$ and $\eta$ poles and using a dispersion
relation for $Re\;\pi^{\pi\eta}(q^2)$, we can write
\begin{equation}
\pi^{\pi\eta}(q^2)\equiv {\theta(q^2)\over (q^2-m^2_\pi+i\epsilon)(q^2
-m^2_\eta + i\epsilon)}\;.
\end{equation}
where the mixing function is related to the mixing angle $\epsilon$ as
\begin{equation}
\theta(q^2) = \left[\epsilon(m^2_\eta) -\epsilon(m^2_\pi)\right] q^2+
\left[m^2_\eta\epsilon (m^2_\pi)-m^2_\pi\epsilon(m^2_\eta)\right]\;.
\end{equation}

Since the $\epsilon$ ``mixing angle" may be a function of $q^2$,
the value of $\epsilon(m^2_\pi)$ may be different from that of
$\epsilon(m^2_\eta)$; in this way the mixing function $\theta(q^2)$
develops a linear
dependence on $q^2$.

On the other hand, in order to study the meson properties from the quark-gluon
degrees of freedom without being plagued by the problem of wave functions,
we need to introduce interpolating quark currents for the correlation
function, and the meson properties are represented by the experimentally
measurable matrix elements of these quark currents.  In the present case,
we choose to work with the axial vector currents rather than pseudoscalar
ones for a number of reasons.  Foremost is that, although $\pi$ and $\eta$
are pseudoscalar particles, the use of the axial vector currents gives a
better convergence property \cite{9,10,11,12}.
Secondly, on the mass shell, the two
currents are directly related.  In an SU(3) notation, we use

\begin{eqnarray}
j^3_\mu(x) & \equiv &{1\over 2}(\bar u\gamma_\mu\gamma_5u(x) -\bar d
\gamma_\mu\gamma_5d(x)) \nonumber \\
j^8_\nu(x) & \equiv & {1\over 2\sqrt3} (\bar u\gamma_\mu\gamma_5u(x) + \bar d
\gamma_\mu\gamma_5 d(x) - 2\bar s\gamma_\mu\gamma_5 s(x))\;.
\end{eqnarray}

The correlator of these two currents is defined by
\begin{equation}
\pi^{\pi\eta}_{\mu\nu}\equiv i\int d^4x\;e^{iqx}<Tj^3_\mu(x)j^8_\nu(0)>\;.
\end{equation}

Using the definition of the mixing angle $\epsilon$, (Eq.~(3)), the PCAC
relation for $\pi$ and $\eta$,\break
\(<0|j^3_\mu(0)|\pi(p)> = i\;f_\pi\;p_\mu,\;<0|
j^8_\mu(0)|\eta(p)> = i\;f_\eta\;p_\mu\),
where $f_\pi$ and $f_\eta$ are the decay constants
for $\pi$ and $\eta$, and the
pole approximation for $Im\pi^{\pi\eta}_{\mu\nu}$, we have:

\begin{equation}
{1\over\pi}\;Im\;\pi^{\pi\eta}_{\mu\nu} = q_\mu q_\nu\;f_\pi\cdot
f_\eta\left[\epsilon
(m^2_\pi)\delta (q^2-m^2_\pi) - \epsilon(m^2_\eta)\delta(q^2-m^2_\eta)\right].
\end{equation}

Because of the structure of PCAC and the $\pi$-$\eta$ pole approximation,
there is no $q^2
g_{\mu\nu}$ term in $Im\;\pi^{\pi\eta}_{\mu\nu}$.  In addition to the $\pi$ and
$\eta$ poles, let us also include the next resonances in the
$Im\;\pi^{\pi\eta}_{\mu\nu}$; these resonances are a pair of pseudo-vector
particles the $b_1$(1235) (denoted by A) and $h_1$(1170) (denoted by B);
these resonances contribute to $Im\;\pi^{\pi\eta}_{\mu\nu}$,

\begin{equation}
(q_\mu q_\nu - q^2 q_{\mu\nu})\tilde f_A\tilde f_B\left[\epsilon'(m^2_A)
\delta(q^2-m^2_A)
-\epsilon(m^2_B)\delta(q^2-m^2_B)\right],
\end{equation}
where $\tilde f_A,\tilde f_B$ are the decay constants for A and B particles.

Combining these two contributions, we can rewrite

\begin{eqnarray}
{1\over\pi}Im\pi^{\pi\eta}_{\mu\nu} & = & q_\mu q_\nu\;Im \pi_1(q^2) - q^2g_
{\mu\nu}\;Im\pi_2(q^2), \nonumber \\
Im\pi_1(q^2) & \equiv & g_\pi\delta(q^2-m^2_\pi) - g_\eta\;\delta(q^2-m^2_\eta)
\nonumber \\
& & +h_A\delta(q^2-m^2_A) - h_B\delta(q^2-m^2_B) \nonumber \\
& & +{\rm higher \ resonances + continuum} \nonumber \\
Im\pi_2(q^2) & = & h_A\delta(q^2-m^2_A) - h_B\delta(q^2-m^2_B) + {\rm higher \
resonances + continuum}
\ ,\end{eqnarray}
where

\begin{eqnarray}
g_\pi & \equiv & f_\pi f_\eta\;\epsilon(m^2_\pi)\qquad\quad
g_\eta\equiv f_\pi f_\eta
\epsilon (m^2_\eta) \nonumber \\
h_A & \equiv & {\tilde f_A\tilde f_B\over m^2_A}\;\epsilon'(m^2_A)\qquad
h_B = {\tilde f_A\tilde f_B\over m^2_B}\;\epsilon'(m^2_B)
\end{eqnarray}

These parameters $g_\pi$, $g_\eta$, $h_A$ and $h_B$
are unknown and will be determined from the QCD sum rules.
Once they are determined in terms of the CSV parameters and the Wilson
coefficients of the OPE for the quark current-current correlation function,
we can substitute these numbers into the mixing function $\theta(q^2)$ and
discuss its momentum dependence in the region of interest.

In the case of $\pi$-$\eta$ mixing, in contrast to the case of $\rho$-$\omega$
mixing, there is no ``on shell" value for $\theta(q^2)$, due to the large
difference of $\pi$ and $\eta$ masses, $\Delta m\equiv m_\eta - m_\pi$ =
412 MeV $\sim 3m_\pi$; this mass difference
presumably comes primarily from the s-quark contribution in the $\eta$.
There is,
thus, no direct experimental method to study $\pi$-$\eta$ mixing, unlike
$e^+e^-\to\pi^+\pi^-$ at the $\omega$-resonance \cite{13,14},
since $\pi$ and $\eta$
cannot be formed from one photon states in $e^+e^-$ collisions; however,
the decay of the $\eta^0$ to $\pi^+\pi^-\pi^0$ via the $\pi^0$ pole serves
to set an ``experimental" mixing at the $\eta$ mass \cite{15}.
Here we shall
content ourselves to use experimentally measured masses for the $\pi$(135),
$\eta$(547), b(1235) and $h_1$(1170) particles as input parameters,
and distinguish between
$\epsilon(m^2_\pi)$, $\epsilon(m^2_\eta)$, $\epsilon'(m^2_A)$, and
$\epsilon'(m^2_B)$ as independent variables.  Although we do not rely on an
expansion in terms of the mass difference $\delta m^2\equiv m^2_\eta -
m^2_\pi$ to
perform the calculations as in the $\rho$-$\omega$ case, the large mass
difference of $\pi$ and $\eta$ is comfortably accommodated without
difficulty.

\section{CALCULATIONS}

One important ingredient of the QCD sum rules is the operator product
expansion (OPE) \cite{16} for the correlation functions.  In our case, we
calculate the OPE of $\pi^{\pi\eta}_{\mu\nu}$ by keeping operators up to
dimension 6.  For the Wilson coefficients, we work to first order in
$\alpha_s$, $\alpha$ and ${mq\over Q}$, where $\alpha_s(\alpha)$ is the fine
structure constant for the strong (electromagnetic) interaction, and $m_q$
is the current mass of quark $q$
($m_u$ for $u$ quark, and $m_d$ for $d$ quark).
We use the definition $Q^2=-q^2$; we do not include the $q^2$ dependence of
$\alpha_s$ but choose its value at 1 GeV$^2$, $\alpha_s$ (1 GeV$^2$)
$\sim$ 0.5.  Our calculation is similar to that for the $\rho$-$\omega$
case but we are dealing with axial vector currents for pseudoscalar mesons;
thus, the reader may convince himself or herself of the signs of our results
by counting the number of $\gamma$ matrices in each diagram.  In the
following, we shall neglect a detailed discussion of each diagram (c.f.
ref.\cite{4})
and simply write down the relevant answers:

\begin{eqnarray}
\pi^{\pi\eta}_{\mu\nu}(q) & \equiv & q_\mu q_\nu \pi_1(q^2) - q^2 g_{\mu
\nu} \pi_2(q^2),\ \ Q^2\equiv -q^2 \nonumber \\
4\sqrt3\pi_1(q^2) & \equiv & C_0\ln\;Q^2 + {C_1\over Q^2} + {C_2\over Q^4} +
{C_3\over Q^6} + O\left({1\over Q^8}\right) \nonumber \\
4\sqrt3\pi_2(q^2) & \equiv & C_0\ln Q^2 + {C_1\over Q^2} - {C_2\over Q^4} +
{C_3\over Q^6} + O \left({1\over Q^8}\right)
\end{eqnarray}
\newpage
where
\begin{eqnarray}
C_0 & = & -{\alpha\over 16\pi^3} + O(\alpha^2_s,\alpha^2),\quad\alpha_s
\equiv {g^2\over 4\pi},\;\alpha\equiv {e^2\over 4\pi} \nonumber \\
C_1 & = & 0 + O(m^2_q) \nonumber \\
C_2 & = & 2[m_u<\bar uu> - m_d<\bar dd>]\nonumber \\
C_3 & = & {352\over 81}\pi\alpha_s[<\bar uu>^2 - <\bar dd>^2] \nonumber \\
    &   & +{88\over 243}\;\pi\alpha[4<\bar uu>^2 - <\bar dd>^2].
\end{eqnarray}
It is helpful to note three points:

\noindent (1) For vector mesons, the conservation of the vector current leads
to a $\pi_{\mu\nu}$ that must be proportional to $q_\mu q_\nu -
q^2g_{\mu\nu}$;
therefore $\pi_1 = \pi_2$.
This is no longer true in the $\pi$-$\eta$ case, since
the axial vector current is not conserved.  A side benefit is that we now have
two independent sum rules for $\pi^{\pi\eta}_{\mu\nu}$, and we can take
advantage of this property to extract information on the parameters appearing
in the mixing function $\theta(q^2)$

\noindent (2) If we restrict ourselves to operators up to dimension 6 and
make the vacuum saturation assumption (VSA) for the four quark condensate
\cite{9}, we find that the $s$ quark in the $\eta$ current (Eq.~(5)) does not
contribute at all.  The reason is that the $s$ quark could only arise in
the four quark condensate $<\bar q\Gamma q\bar s\Gamma s>$ ($q$ = up or down
quark) which vanishes under the VSA.

\noindent (3) Taking the difference of the sum rules for $\pi_1$ and $\pi_2$
we see that the coefficient $C_3$ for the dimension 6,
four-quark condensate drops out.
Therefore the uncertainty in this coefficient due to the assumption of the VSA
can be ignored.

The next step is to relate these CSV parameters and condensates to the
phenomenological parameters $g_\pi$, $g_\eta$, $h_A$ and $h_B$ in the RHS of
$Im\;\pi_{\mu\nu}$ (our model for $Im\;\pi_{\mu\nu}$).  In order to assure
convergence, to accentuate the lower resonances and to lessen the effect of
the continuum, we follow the standard practice of applying a Borel
transformation to $\pi_1$ and $\pi_2$ \cite{9},

\begin{equation}
\hat L_M[f(Q^2)]\equiv\lim_{{{{n\to\infty\atop
Q^2\to\infty}\atop{{Q^2\over
n} = M^2}}\atop {\rm fixed}}}\;{1\over (n-1)!}(Q^2)^n\left(-{d\over dQ^2}
\right)^n\;.
\end{equation}
In this way, we obtain a set of simultaneously linear equations for $g_\pi$,
$g_\eta$, $h_A$ and $h_B$:
\newpage
The sum rule for $\pi_1$ gives
\begin{eqnarray}
-C_0+C_2\left({1\over M^4}\right) & + & {1\over 2}C_3\left({1\over M^6}\right)
\nonumber \\
& = & {4\sqrt3\over M^2}\left[g_\pi\;e^{-{m^2_\pi\over
M^2}}-g_\eta\;e^{-{m^2_\eta\over M^2}} + h_A\;e^{-{m^2_A\over M^2}} -h_B\;
e^{-{M^2_B\over M^2}}\right] \nonumber \\
&  & +\;{\rm higher \ resonances} (\pi',\eta')(A',B') + {\rm continuum}\;,
\end{eqnarray}
and that for $\pi_2$ yields
\begin{eqnarray}
-C_0 - C_2\left({1\over M^4}\right) & +
 & {C_3\over 2}\left({1\over M^6}\right)
\nonumber \\
& = & {4\sqrt3\over M^2} \left[h_A \ e^{-{m^2_A\over M^2}}
- h_B \ e^{-{m^2_B\over
M^2}}\right] + {\rm higher \ resonances} (A',B') + {\rm continuum}\;.
\end{eqnarray}

Taking the difference, we get
\begin{equation}
2C_2\left({1\over M^4}\right) = {4\sqrt3\over M^2}\left[g_\pi
\ e^{-{m^2_\pi\over
M^2}} - g_\eta \ e^{-{m^2_\eta\over M^2}}\right] + {\rm higher \ resonances}
(\pi',\eta')\;.
\end{equation}
It is worthwhile to call attention to the fact that the contributions of higher
resonances $b_1$(1235) and $h_1$(1170) and continuum have been completely
canceled in Eq. (16).  The same is true for the electromagnetic continuum,
whereas higher resonances in the pseudoscalar channel, such as $\pi'$(1300)
- $\eta'$(1290), will survive (in principle) in Eq.~(16).  However, it will
turn out that taking these resonances explicitly into account does not change
the Borel analysis noticeably.  This is in contrast to the case of
$\rho$-$\omega$ mixing where the corresponding high resonances
($\rho'$-$\omega'$) are important.

By taking the derivative with respect to
$\left({1\over M^2}\right)$ on both sides
of Eq. (16), we obtain another sum rule:

\begin{equation}
{-C_2\over 2\sqrt3} = m^2_\pi \ g_\pi \ e^{-{m^2_\pi\over M^2}} -
m^2_\eta \ g_\eta \
e^{-{m^2_\eta\over M^2}}\;.
\end{equation}

With these two equations, we can solve for $g_\pi$ and $g_\eta$:

\begin{eqnarray}
g_\pi & = & \left({C_2\over 2\sqrt3 M^2}\right) \ e^{m^2_\pi/M^2}\left({M^2+m^2
_\eta\over m^2_\eta - m^2_\pi}\right) \nonumber \\
g_\eta & = & \left({C_2\over 2\sqrt3 M^2}\right) \ e^{m^2_\eta/M^2}\left({M^2+
m^2_\pi\over m^2_\eta - m^2_\pi}\right)
\end{eqnarray}

Finally, the mixing function is
\begin{equation}
\theta(q^2)\equiv \alpha(M^2)q^2+\beta(M^2)\;,
\end{equation}
where
\begin{eqnarray}
\alpha(M^2) & \equiv & {g_\eta-g_\pi\over f^2_\pi}\;, \nonumber \\
\beta(M^2) & \equiv & {m^2_\eta g_\pi - m^2_\pi g_\eta\over f^2_\pi}
\;.
\end{eqnarray}
where we assume the SU(3) symmetry for the decay constants $f_\pi = f_\eta$.

In Fig.~1 we plot the dependence of $g_\pi$ and $g_\eta$ on $M^2$.
In the preferred region of 1.0 GeV$^2\lesssim M\lesssim 1.6$ GeV$^2$,
the difference
between $g_\pi$ and $g_\eta$ is small.  Furthermore, it can be seen from
Fig.~1 and Eq.~(18) that if $M^2\gg m^2_\eta$, $g_\pi\approx g_\eta$.  We thus
find that $\theta(q^2)$ has a weak $q^2$ dependence.

In order to fix the values of $g_\pi$, $g_\eta$, $\alpha$ and $\beta$,
we choose a suitable
Borel window and average the functions $g_\pi(M^2)$, $g_\eta(M^2)$ over the
range of $M^2$ (1 GeV$^2$ $\sim$ 1.5 GeV$^2$).  At this point, one should
note that if we included the higher resonances
$\pi$(1300) and $\eta$(1295) in the phenomenological side of the sum rules
and left their masses as parameters, we could perform the stability
analysis to obtain the ``optimal" values for $g_\pi$ and $g_\eta$.  However,
our numerical results do not suggest such ``optimal" values are well-defined.
The numerical values we obtain
for these parameters are

\begin{equation}
\begin{array}{ll}
g_\pi = 114 \ {\rm MeV}^2 & g_\eta = 117 \ {\rm MeV}^2 \nonumber \\
 \alpha = 3.65\times 10^{-4} & \quad\beta = 3660 \ {\rm MeV}^2
\end{array}
\end{equation}
All numbers have an error of about 10-20\%, which is due to the uncertainty
of the isospin symmetry violating quantities ${m_d-m_u\over m_d+m_u}$ and
$\gamma ={<\bar dd>\over <\bar uu>}-1$ in the coefficient $C_2$ and the fact
that $M^2\gtrsim 0.9$ GeV$^2$ for stability.

\section{COMPARISONS WITH OTHER MODELS}

In this section, we shall compare our results with those of other authors,
who use a different approach.

$\pi$-$\eta$ mixing has been considered previously by
Maltman and Goldman (MG) \cite{5,6} and Piekarewicz (P) \cite{7}.
Maltman used chiral
perturbation theory \cite{5} to calculate
the $\pi$-$\eta$ mixing to the one-loop
order.  The mixing function $\theta(q^2)$ is expressed in terms of meson
masses and other physical observables; theoretical uncertainty comes
from the electromagnetic mass difference of kaons.  In order to verify the
assumption which underlies the GHT calculation.
Maltman and Goldman \cite{6} also
use a
chiral quark model to calculate $\pi$-$\eta$ mixing and get a consistent
result with chiral perturbation theory.

Piekarewicz \cite{7} used a purely hadronic model to calculate
$\pi$-$\eta$ mixing, where the mixing is generated by a $N\bar N$ loop.  In
order to fix the renormalization point for the divergent integral, he chose
the ``on shell" value of $m_\eta$ to fix the intercept of the mixing function.

He also
derived a ratio between the slope at the origin of the $\rho$-$\omega$
and $\pi$-$\eta$ mixing amplitudes
\begin{equation}
{<\pi|H|\eta> \ {\rm slope \ at} \ q^2=0\over <\rho|H|\omega> \ {\rm slope \
at} \ q^2=0}\simeq {g_\pi g_\eta\over f_\rho g_\omega}
\end{equation}
which he claimed to be less model-dependent and to hold approximately
for a reasonable range of $q^2$.

In view of the lack of ``on-shell" experimental value for the $\pi$-$\eta$
mixing amplitude (the previous calculations are based on SU(3) mass splitting
\cite{17,18} of pseudoscalar mesons and a pole
model analysis of the $\eta$-$\eta'$
system \cite{17,18}), our calculation provides an independent result for the
slope and intercept for the mixing function.  We list the different results
from the three calculations in Table 1.

It is to be noticed that our slopes are three times smaller than that of
chiral perturbation theory
and correspondingly so is the ratio $\alpha/\beta$.
Because of the small slope, our
mixing function only changes about
5\% from $q^2 = m^2_\eta$ to $q^2 = -m^2_\eta$.
Therefore, the mixing function is practically a constant.  On the other hand,
the intercept of our mixing function ($\beta$) is fairly close to that obtained
in chiral perturbation theory.  Furthermore it should be noted that both in
the GHT and in the PW approaches the $q^2$ variation of the $\pi$-$\eta$
mixing amplitude is much smaller than the one for the $\rho$-$\omega$
mixing amplitude obtained in the same approaches \cite{1,2,3}.
Finally, our mixing function gives an
``on shell" value $(q^2 = m^2_\eta)$ of 3800 MeV$^2$, which is slightly
smaller (but within our estimated error) than that of the other two
approaches and the one of 4200 MeV$^2$
obtained from the CSB NN force \cite{17,18,19}.

\section{SUMMARY AND CONCLUSIONS}

In this paper, we use QCD sum rules to study $\pi$-$\eta$ mixing,
choosing a combination of the two sum rules coming from the correlation
functions of two axial vector currents.  We are able to extract the leading
behavior of the mixing function $\theta$ as a function of $q^2$ without
the need to use four-quark condensates.  We obtain $\theta(q^2=m^2_n) =
3800$ MeV$^2$, and a $q^2$-dependence of $\theta(q^2)$ that is compatible
with zero.  This is in contrast to the $\rho$-$\omega$ mixing amplitude,
which varies strongly with $q^2$.  Our results are in qualitative agreement
with those obtained from various quark and hadronic models as well as from
chiral perturbation theory.

\begin{figure}
\caption{The mixing parameters $g_\pi$ and $g_\eta$ as functions of the
squared Borel mass $M^2$}
\end{figure}

\begin{table}
\caption{Comparisons of different approaches for the $\pi$-$\eta$ mixing
function}
\begin{tabular} {lcccc}
& {$\alpha$} & $\beta$ (MeV$^2$) & $\alpha/\beta$ (MeV$^{-2}$) &
$\theta(q^2 = m^2_\eta$)(MeV$^2$) \\
\tableline
Chiral Perturbation & 1.08 $\times 10^{-3}$ & 3808 & 2.836 $\times 10^{-7}$
& 4131.68 \\
Theory \cite{4,20} & & & & \\
\tableline
Hadronic Model \cite{6} &1.7$\times$ 10$^{-3}$ &3800 & 4.4$\times$ 10$^{-7}$
&4200 \\
\tableline
QCD Sum Rules & 3.5$\times 10^{-4}$ & 3700 & 9.45 $\times 10^{-8}$ & 3800 \\
\end{tabular}
\end{table}

\end{document}